\documentclass[aps,twocolumn]{revtex4}

\usepackage{epsfig} 
\usepackage{graphics}

\newcommand{\ket}[1]{| #1 \rangle}

\begin{document}
\title{Spin squeezing of atoms by Rydberg blockade}
\author{Isabelle Bouchoule and Klaus M{\o}lmer}
\affiliation{
Institute of Physics and Astronomy
University of Aarhus\\
DK 8000 Aarhus C., Denmark
}

\begin{abstract}
We show that the interaction between Rydberg atomic states 
can provide continuous spin squeezing of atoms with two ground states.
The interaction prevents the simultaneous excitation of more than 
a single atom in the sample to the Rydberg state, and
we propose to utilize this blockade effect to realize an effective 
collective spin hamiltonian $J_x^2-J_y^2$. 
With this hamiltonian the quantum mechanical
uncertainty of the spin variable
$J_x+J_y$ can be reduced significantly. 
\end{abstract}

\maketitle

\vspace{0.5cm}

The sizable dipole interaction between atoms which have been transferred
with pulsed laser fields to highly excited Rydberg states has been
proposed [1,2] as a mechanism for entanglement operation on the state
of neutral atoms. A "Rydberg blockade" effect realized by the dipole
interaction 
prevents more than one atom to enter a Rydberg state at a time.
Hence, the evolution of one atom can be conditioned on the state
of another one as requested for a two-qubit gate in a quantum computer
\cite{Jaks00}. The Rydberg blockade effect can also be used to 
change the state of an entire atomic ensemble, one atom at a time,
and can potentially be used to produce a single atom 'gun' \cite{Luki00}.

 In this letter, we show that a combination of the Rydberg blockade effect
and continuous  laser fields can produce  particular entangled states of 
an ensemble of atoms, the so-called  spin squeezed states. 
Spin squeezing refers to the collective spin 
$\vec{J} = \sum_i\vec{S}_i$ of a collection of spin 1/2 particles, for which 
the Heisenberg inequality 
 assures $\Delta {J}_x \Delta {J}_y \geq |\langle
{J}_z\rangle|/2,\ (\hbar=1)$. 
A state whose mean spin is along $z$ and 
in which the width of the distribution of $J_x$
is reduced so that $\Delta {J}_x < \sqrt{|\langle {J}_z\rangle|/2}$
is called spin squeezed.
The spin notation symbolises 
the state of an ensemble of two-level atoms, where
the two states are represented as the components $\pm 1/2$ along $z$ of
a spin 1/2 particle, and
spin squeezing is a useful property
since reduced spin fluctuations imply an improvement of the 
counting statistics for the number of atoms in
specific states, {\it i.e.}, improved resolution in spectroscopy and
in atomic clocks \cite{Wine94,Sant99}. 

Recently, a number of proposals for spin squeezing and atomic
noise reduction has been made
involving absorption of broad band  squeezed light \cite{Kuzm97,Hald99},
collisional interactions in two-component condensates
\cite{Sore01,Poul01}, and quantum non-demolition detection of 
atomic populations \cite{kuzm00,kuzm98,Molm99_EurJPhys}.
In the work presented here, an atomic gas is illuminated with lasers which
couple long lived
states $\ket{a}$ and $\ket{b}$ 
to a Rydberg state $\ket{r}$. The lasers are far detuned
so that the the population in the Rydberg state is small and their
effect is described  by an effective hamiltonian $H$ acting on the 
states $\ket{a}$ and $\ket{b}$. 
We first show how non-linearities appear in
the simple case of the lightshift produced by a single laser.
The hamiltonian $J_z^2$ is realised and squeezing will occur. 
This hamiltonian, however, has the drawback that the squeezing axis 
depends on the interaction time and on the total number of atoms. 
Thus, we propose a way to realise the hamiltonian 
$J_x^2-J_y^2$ which  enables stronger squeezing and which also presents the
advantage that the squeezing axis is stationary \cite{Kita93}.

\begin{figure}
\centerline{\includegraphics{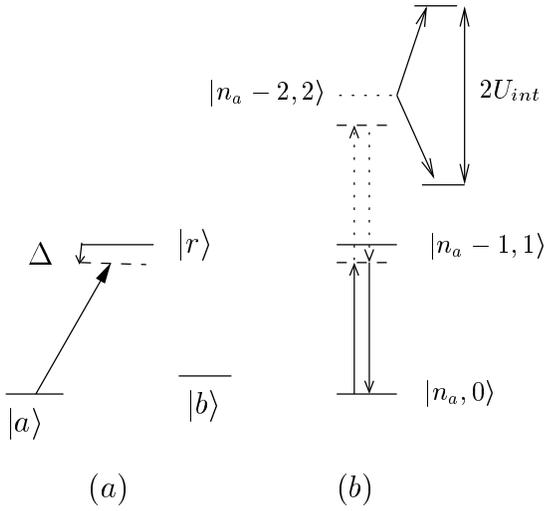}}
\caption{\it Laser configuration and relevant states for calculation of 
the light shift to fourth order in the presence of a single laser.
(a) Energy levels of a single atom. (b) Energy levels of a collection
of atoms: the upper part of the figure shows how interaction causes
an upward or downward shift $U_{int}$ of the state with 
two Rydberg excited atoms.}
\label{fig.1laser}
\end{figure}

Let us consider the situation depicted in Fig. \ref{fig.1laser}, 
where an ensemble of $N$ atoms is illuminated by a laser field detuned
by $\Delta$ from resonance of the transition $\ket{a}\rightarrow \ket{r}$.
If the internal state of the atoms is initially  symmetric with 
respect to exchange of atoms, we can consider only the symetric states 
and a basis is formed by the states $\ket{n_a,n_r}$, where $n_a$ is the 
number of atoms in the state $\ket{a}$, $n_r$ is the number of atoms in 
the state $\ket{r}$, and the remanining $N-n_a-n_r$ atoms populate the
state $|b\rangle$. 
If the laser is sufficiently weak, the population in the state with $n_r
> 0$ is very small, and the only effect of the laser is to
shift the energy of the states $\ket{n_a,0}$.
The state $\ket{n_a,0}$ is coupled with the amplitude $\sqrt{n_a}\Omega$
to $\ket{n_a-1,1}$ which, in turn, is coupled to the state
$\ket{n_a-2,2}$ with the amplitude $\sqrt{2}\sqrt{n_a-1}\Omega$.
The expression of the light shift to fourth order in the laser field
amplitude is therefore
\begin{equation}
\Delta E_{n_a}=-n_a\frac{\Omega^2}{\Delta}
             +n_a^2\frac{\Omega^4}{\Delta^3}
             -\frac{1}{2\Delta}\frac{2 n_a(n_a-1) \Omega^4}{\Delta^2},
\label{Ls4.eq}
\end{equation}
where the last term is due to a two photon transition to the state 
$\ket{n_a-2,2}$.
 The terms proportional to $n_a^2$ in $\Delta E_{n_a}$ cancel and the light 
shift is proportional to $n_a$ as expected for non interacting atoms. Indeed,
the energy of the state $\ket{a}$ of each atom does not depend on the state
of the other atoms.  In the picture suggested by Fig.\ref{fig.1laser}.(b), 
the absence of non-linearities for non interacting atoms is due 
to destructive interference
between processes involving states with at most one atom in the 
Rydberg state and processes involving states with several atoms in 
the Rydberg state.

Let us now assume that atoms in the Rydberg state interact so that the 
energy of the states $\ket{n_a-2,2}$ is shifted by $\pm U_{int} \gg \Delta$.
Then, the two photon contribution to the light shift 
is negligible and the light shift of $\ket{n_a,0}$ is given by the two
first terms of Eq.(1):
\begin{equation}
\Delta E_{n_a}=-n_a\frac{\Omega^2}{\Delta}
             +n_a^2\frac{\Omega^4}{\Delta^3}.
\label{eq.LS4int}
\end{equation}    
By removing interference paths with
more than one atom in the Rydberg state,
the``Rydberg Blockade''  leads to the non-linear interaction.
Note that the light shift (2) is independent of the precise 
interaction strength
between Rydberg excited atoms. This implies that as long as the 
interaction is strong enough to substantially increase the detuning, 
{\it i.e.} for atoms with a wide range of spatial separations,
the light shift  is   given by
Eq.(\ref{eq.LS4int}).   
 Writing $n_a=J_z+N/2$, we see that the quadratic light shift in 
$n_a$ results in an effective  hamiltonian containing a term in $J_z^2$.
Such a hamiltonian, applied to 
an initial coherent spin state directed in the (x,y) plane, 
gives squeezing\cite{Kita93}. The terms linear in $J_z$ in the hamiltonian are
responsible for a rotation of the spin. The addition of a second laser,
affecting the atomic state $\ket{b}$, enables us to realise a rotation
independent of the number of atoms \cite{Bouc01_2}.  

A better hamiltonian to produce squeezing is 
\begin{equation}
H=2\Omega_{\rm eff}\left ( J_x^2-J_y^2\right )=
\Omega_{\rm eff}\left ( a^2{b^+}^2 + {b}^2{a^+}^2 \right ).
\label{eq.a2b2}
\end{equation}
 It corresponds to the transfer of atoms to $\ket{b}$ in pairs, and 
it is thus analogous to the hamiltonian for production of squeezed light
which creates and annihilates photons in pairs.
 If this hamiltonian is applied to an ensemble of atoms initially 
in $\ket{a}$, the spin variance 
$\langle J_{-\pi/4}^2\rangle=
\langle (e^{i\pi/4}a^+ b +e^{-i\pi/4}b^+ a)^2\rangle$ is reduced.
Furthermore, the best squeezing we can achieve with 
such a hamiltonian 
is larger than the one we can achieve with $J_z^2$.
We propose to realize the hamiltonian (3) in the following way.

\begin{figure}
\centerline{\includegraphics{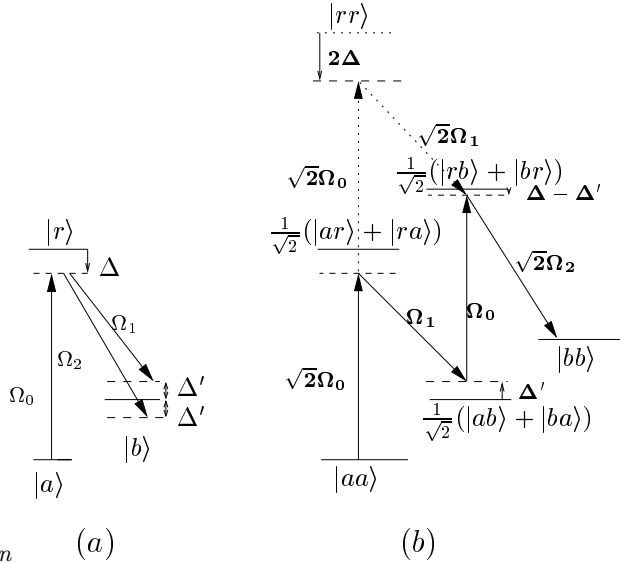}}
\caption{\it (a) Energy levels in a single atom and 
transitions induced by laser fields
$\Omega_i,\ i=0,1,2$ to couple the 'spin' states $|a\rangle$ and
$|b\rangle$ via the intermediate Rydberg state $|r\rangle$.
(b) Transition paths transfering two atoms from the state $\ket{aa}$
to the state $\ket{bb}$. The first path (solid lines) does not use the 
state $\ket{rr}$, the second path (dotted lines) does. 
If the atoms interact in  the  state $\ket{rr}$, so that this level
is shifted by an amount much
larger than $\Delta$, the amplitude of the dotted path becomes negligible,
and a net coupling appears from $\ket{aa}$ to $\ket{bb}$. 
} 
\label{fig.paths}
\end{figure}

 As shown Fig.\ref{fig.paths}.(a), 
Raman couplings between $\ket{a}$ and $\ket{b}$
are introduced by three laser fields with
two Stoke  fields, $\Omega_1$ and $\Omega_2$,
detuned symmetrically around the Raman resonance by
the amount $\pm \Delta'$. The idea is now that a single atom will not
make the transition between states $|a\rangle$ and $|b\rangle$ because
it is not resonant, but {\it two} atoms can simultaneously make
the transition $|aa\rangle \leftrightarrow |bb\rangle$ since this
process occurs resonantly if one atom emits a Stokes photon stimulated
by $\Omega_1$ and the other emits a photon stimulated by $\Omega_2$.

Consider two atoms initially in the product state $\ket{aa}$ 
illuminated by lasers with equal couplings for both atoms as 
depicted in Fig.\ref{fig.paths}.(b).
If the atoms do not interact, 
there is no way they can exchange the energy mismatch of the
stimulated emissions in the fields $\Omega_1$ and $\Omega_2$
and the effective coupling to $\ket{bb}$ vanishes. 
As in the previous proposal, this can be understood 
in terms of the destructive interferences between 
paths involving the state where both atoms are in $\ket{r}$
($\ket{rr}$) and the other paths. Fig. \ref{fig.paths}.(b)
shows the paths for which stimulated emission 
in the field $\Omega_1$ occurs first.

In contrast, if the interaction between the atoms shifts the
energy of the state $\ket{rr}$ 
by $\pm U_{int}\gg \Delta$, the amplitude of the paths involving the
intermediate state $\ket{rr}$ (dotted line in 
Fig.\ref{fig.paths}.(b)) is suppressed
compared to the paths represented by the solid line in the figure, 
the destructive interference is suppressed, and 
the coupling between $\ket{aa}$ and $\ket{bb}$ is now
\begin{equation}
\Omega_{c}=-\frac{4\Omega_0^2\Omega_1\Omega_2}
{\Delta(\Delta-\Delta')(\Delta+\Delta')}.
\label{eq.omegaeff}
\end{equation} 
A similar four photon transition 
has been used to entangle ions in
ion traps \cite{Sore99ions,Sack00}, where the suppression
of destructive interference arises from the Coulomb interaction
which lifts the  degeneracy of collective vibrational modes.
We note that the emergence of a resonant transition due to
removal of interfering transition paths also has analogies in
spectrocopy on gasses, where different mechanisms for pressure induced
resonances work by similar mechanims \cite{Vara92,Gryn90}.

As $U_{\rm int}$ scales as $1/r^3$, 
the coupling between $\ket{aa}$ and $\ket{bb}$ is given by 
(\ref{eq.omegaeff}) as long as the distance between the atoms is smaller than 
a given critical distance $d_0$ for which $U_{\rm int}\gg \Delta$.
 Thus, in an atomic sample with a size  smaller than $d_0$ the
transfer of atoms from $\ket{a}$ to $\ket{b}$ is represented by
the squeezing hamiltonian (\ref{eq.a2b2}) with 
$\Omega_{\rm eff}=\Omega_{c}/2$. Terms involving more than two 
atoms at a time would be of higher order in the Rabi frequencies of the 
lasers and are neglected.  

We now turn to an analysis of the time required to obtain substantial
spin squeezing.  The coupling between states with $n_b$ and
$n_b+2$ atoms transfered to $\ket{b}$ is about the same as the one 
between harmonic oscillator number states introduced by the 
squeezing hamiltonian  $N\Omega_{\rm eff}(b^2+{b^+}^2)$,
as long as $n_b$ is much smaller
than the total number of atoms $N$.
Thus, we expect the squeezing to evolve as
\begin{equation}
\left \langle J_{-\pi/4}^2  \right \rangle(t)=
e^{-4N\Omega_{\rm eff} t} \left \langle J_{-\pi/4}^2  \right \rangle(0)
\label{eq.squeezing}
\end{equation}
and the mean number of atoms in $|b\rangle$ to follow
\begin{equation}
 \overline{n_b}=\sinh^2(2N\Omega_{\rm eff} t).
\label{eq.nb}
\end{equation} 
For ease of presentation we introduce the amount of squeezing, 
$S=(N/4)/\langle J_{-\pi/4}^2\rangle$. 
For intermediate times so that $1\ll \overline{n_b}\ll N$,
the squeezing verifies
$S\simeq 4\overline{n_b}$. 
Solving numerically the evolution produced by the hamiltonian 
(\ref{eq.a2b2}), we find that these simple analytical expressions are 
accurate up to 5\% as long as $\overline{n_b}<0.05 N$ and that the maximum 
squeezing obtained is about $S \simeq N/2$.

The  amplitudes for the excitations of Rydberg states from state
$|a\rangle$ and state $|b\rangle$ are
proportional to $\sqrt{n_a}\Omega_0$ and $\sqrt{n_b}\Omega_{1,2}$, 
respectively.
To justify the elimination of the Rydberg state,
the coupling amplitudes should therefore obey
\begin{equation}
\left \{
\begin{array}{l}
\sqrt{N}\,\Omega_0  \ll \Delta\\
\sqrt{S/4}\,\Omega_1  \ll \Delta+\Delta'\\
\sqrt{S/4}\,\Omega_2  \ll \Delta-\Delta'\\
\end{array}
\right .
\label{eq.adia}
\end{equation}
Here, we took $S\simeq 4n_b$.
Assuming  that these inequalities  are all fulfilled by an order
of magnitude, and taking $\Delta\pm \Delta'\sim \Delta$,
the time required to obtained the squeezing $S$ is about
\begin{equation}
T \simeq \frac{1}{16}\frac{10^4}{\Delta}S\ln(S). 
\label{eq.tempssqu}
\end{equation} 
$T$ is almost
linear in the squeezing parameter $S$,  and  does not depend  on the 
total number of atoms.
$T$ should be short enough so that incoherent effects
such as spontaneous emission or thermal field absorption 
do not alter  the squeezing significantly. 
To estimate the effect of such incoherent processes, we consider the 
simple case of loss of atoms, which represents, for example, 
atoms decaying to a ground state different from $\ket{a}$ and 
$\ket{b}$ after spontaneous emission.  If 
the atom $i$ has been lost, the spin variance of the remaining atoms is
$\langle J_x'^2  \rangle=\langle (J_x -S_{x_i})^2 \rangle$ where $S_{x_i}$ 
is the spin of the lost particle. Due the the permutation symmetry of
the atomic state, $\langle \sum_{j}S_{x_j}S_{x_i}\rangle=
\frac{1}{N} \langle J_x^2\rangle$, and 
we  get
\begin{equation}
\langle J_x'^2  \rangle=\langle J_x^2  \rangle\left ( 1-\frac{2}{N}\right )
+\frac{1}{4}.
\end{equation}
After the loss of $n_L$ atoms, we thus have the reduced squeezing
\begin{equation}
S'=\frac{(N-n_L)/4}{\langle J_x'^2  \rangle} \simeq 
S\cdot \frac{1}{1  +\frac{n_LS}{N}}
\label{eq.varSloss}
\end{equation} 
where $S$ is the value of the squeezing before the losses.
To have a negligible effect of losses on the squeezing, 
we require $n_LS/N\ll 1$.
The sensiblility of squeezing to losses increases as the 
squeezing increases. This behavior is expected as a strong squeezing
corresponds to strong 
correlations and entanglement of the particles \cite{Sore01_Maxsqueeze}.  
With a population of the Rydberg state of about $10^{-2}$, which 
correspond to the inequalities (\ref{eq.adia}) fulfilled by a factor 10,
the expression (\ref{eq.tempssqu}) for the squeezing time implies that
the atom decay or loss rate $\Gamma$ should obey
\begin{equation}
  \frac{\Delta}{\Gamma}\gg \frac{10^2\,S\ln(S)}{4} 
\label{eq.MinDelta}
\end{equation}
For Rydberg atoms with $n\sim 50$, an overestimate for the 
spontaneous emission rate and the interaction
with the black body field yields $\Gamma \sim 10\,$kHz\cite{Luki00}. Thus,
to obtain a squeezing of a factor 10,  we require $\Delta > 6\,$MHz.
For $\Delta=50\,$MHz 
the interaction energy between the Rydberg atom is much larger 
than $\Delta$ for a distance between atoms 
$d \leq 3\mu $m\cite{Luki00}. With a density of atoms of
$2\times 10^{11}$at/cm$^3$,
realised in atomic ensembles obtained from a magneto-optical trap,
each atom interacts with about 20 other atoms. Thus, a squeezing 
by a factor about 10 can be obtained. 

\begin{figure}
\centerline{\includegraphics{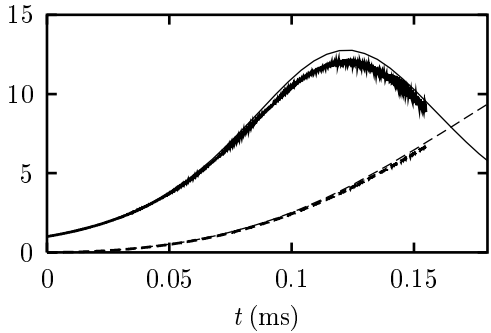}}
\caption{\it Evolution of the squeezing factor $S=(N/4)/<J_{-\pi/4}^2>$ as 
a function of time. Thick lines : Numerical evolution with  
6 lasers as explained in the text with $\Delta=50\,$MHz,
$\Delta'=20\,$MHz and $|\omega_0|=|\omega_1|=|\omega_2|=1.1$\,MHz.
Thin lines : Evolution according to the hamiltonian (\ref{eq.a2b2}). The
dashed lines show the evolution of the number of atoms 
in the state $\ket{b}$ (\ref{eq.a2b2}).}
\label{fig.evol}
\end{figure}

The coupling introduced by the lasers 
is well represented by the hamiltonian (\ref{eq.a2b2}),
but as seen in the previous section,
the dipole Blockade effect is accompanied by a non linearity in 
the lightshift of the states and the resulting nonlinear terms
inhibit the evolution towards states with significant
squeezing.  To cancel these lightshifts, we propose to add three 
other lasers of the same intensity but with opposite  detuning of the 
laser fields indicated in Fig.\ref{fig.paths}.(a). These lasers contribute
to both the lighshift and the two-atom Raman coupling. 
If the two added Stokes field are dephased respectively by
$+90$ and $-90$ degrees with the original ones, the net effect
is a vanishing lightshift but a non vanishing Raman coupling.
Figure \ref{fig.evol} shows the calculated evolution of 20 atoms illuminated 
by six lasers with appropriate relative phases. 
Only states with $n_r<2$ have been 
taken into account in the calculation since, 
in the presence of a strong interaction
between Rydberg atoms, they are the only states relevant.
The numerical 
results follow the results of the simple quadratic spin
hamiltonian (\ref{eq.a2b2})  
with a small discrepancy due to even higher order terms in the lightshift.
The maximum squeezing factor is  within a factor 2 equal to the total 
number of interacting atoms.

In summary, we have proposed a mechanism to produce spin squeezed states
of atoms by use of a Rydberg blockade effect induced by {\it cw} laser
fields. Note that
many other interaction mechanisms may produce
a similar blockade of destructive interference. Due to
the interference blockade, bichromatically driven 
quantum transitions  via  intemediate states with enhanced 
interparticle interactions, will in general lead to pair-wise transitions
and non-linear collective dynamics of the ensemble. For the Rydberg
blockade mechanism we have analyzed 
 the allowed range of parameters, and our
calculations show that reduction of the collective spin noise
by a factor about 10 is possible with current experimental 
parameters.
These results were obtained for the case of an ensemble of size smaller
than $d_0$ so that the interaction energy between any two atoms 
in the Rydberg state is sufficient to ensure the blockade effect,
$U_{\rm int} \gg \Delta$.
It is experimentally relevant to analyze the alternative 
case of 
\pagebreak

a spatially large sample, where the Rydberg blockade is
effective only for the $n$ nearest neighbours of a given atom, $n \ll N$. 
In that case each atom gets entangled with atoms in its neighbourhood
which in turn gets entangled with atoms further away, etc.
Simulations of such a coupling show that to a good approximation,
the squeezing evolves similar to that
in an entire ensemble with $n$ atoms all interacting with 
each other\cite{Bouc01_2}. 
The expected amount of squeezing is thus on the order of $n/2$ 
and, squeezing factors around $S=10$ should be realistic in
macrosopic samples. If the atoms move around during the illumination, 
as in a gas, they are in a time integrated sense brought into contact
with a larger number of other atoms. If their speeds are large
enough to constantly introduce new interacting pairs,
{\it i.e.}, if $v_{rms}T >> d_0$, the squeezing factor
will continue to grow as $e^{4n\Omega_{eff}t}$ beyond the 
maximum allowed for $n$ atoms. 

Isabelle Bouchoule thanks the European Community for her Marie-Curie 
Fellowship.


\end{document}